\documentclass[runningheads]{llncs}
\usepackage{graphicx}
\usepackage{booktabs}
\usepackage[table]{xcolor}
\begin{document}
\title{Ultrasound Liver Fibrosis Diagnosis using Multi-indicator guided Deep Neural Networks}
%
%
\newcommand\blfootnote[1]{%
\begingroup 
\renewcommand\thefootnote{*}\footnotetext{#1}%
\addtocounter{footnote}{0}%
\endgroup 
}

\newcommand\bifootnote[1]{%
\begingroup 
\renewcommand\thefootnote{$\dag$}\footnotetext{#1}%
\addtocounter{footnote}{0}%
\endgroup 
}

\author{Jiali Liu \inst{*12} \and
Wenxuan Wang \inst{*12} \and
Tianyao Guan\inst{2} \and Ningbo Zhao\inst{3} \and  Xiaoguang Han\inst{12} \and Zhen Li   \inst{12}}
\authorrunning{Liu et al.}
\institute{Shenzhen Research Institute of Big Data. Shenzhen, Guangdong , China
 \and Chinese University of Hong Kong, Shenzhen, Shenzhen, Guangdong, China   \and
The Third People’s Hospital of Shenzhen , Shenzhen, China}

\blfootnote{denotes equal contribution, in alphabetical order}
\bifootnote{denotes corresponding author: lizhen@cuhk.edu.cn}

\maketitle 

\begin{abstract}
Accurate analysis of the fibrosis stage plays very important roles in follow-up of patients with chronic hepatitis B infection. In this paper, a deep learning framework is presented for automatically liver fibrosis prediction. On contrary of previous works, our approach can take use of the information provided by multiple ultrasound images. An indicator-guided learning mechanism is further proposed to ease the training of the proposed model. This follows the workflow of clinical diagnosis and make the prediction procedure interpretable. To support the training, a dataset is well-collected which contains the ultrasound videos/images, indicators and labels of 229 patients. As demonstrated in the experimental results, our proposed model shows its effectiveness by achieving the state-of-the-art performance, specifically, the accuracy is \textbf{65.6\%}  (\textbf{20\%} higher than previous best).

\end{abstract}
\section{Introduction}
It was estimated that about 248 million people worldwide were chronic HBV infections in 2010~\cite{1schweitzer2015estimations}. Most people with HBV infection would develop to cirrhosis or hepatocellular carcinoma. As reported, the number of deaths from cirrhosis and or hepatocellular carcinoma caused by HBV increased by 33\% between 1990 and 2013~\cite{2stanaway2016global}.
An accurate analysis of the fibrosis stage is thus very important during follow-up of patients with chronic hepatitis B infection. The development of fibrosis is not only a major prognostic factor, but often used to determine whether a patient needs antiviral therapy. Although liver biopsy is considered of the gold standard, it has many shortcomings. The most important thing is that the false negative result for diagnosing cirrhosis can reach 20\% to 30\% due to sampling errors~\cite{3denzer2007prospective}. As a result, patients will usually miss the optimal treatment time. In addition, biopsy is invasive and expensive~\cite{4wong2000watchful}, which brings great pain and a heavy financial burden to the patient. There is also a risk of serious complications (0.4\%) and may even result in death (0.03\%). Therefore, various non-invasive methods have been developed to replace the role of liver biopsy in fibrosis staging~\cite{5dohan2014transjugular}, such as measuring related biomarkers and observing their morphological changes by ultrasound or magnetic resonance.
Considering ultrasonic examination is non-radiation, cheap and easy-to-access in practice, this paper focuses on designing algorithms for automatic liver fibrosis diagnosis using ultrasound images. 


Previously, many works have utilized conventional machine learning methods for this task. The work of ~\cite{8mojsilovic1998characterization} developed a method for textured feature modeling, which is then applied to classify liver disease. Yeh et al.~\cite{9yeh2003liver} proposed to classify the liver fibrosis status using the features extracted by gray level concurrence and non-separable wavelet transform. Advanced deep learning approaches, such as convolutional neural networks (CNN), show high superiority than conventional learning methods on the feature modeling ~\cite{10simonyan2014very}. They are also successfully applied in some applications of medical image analysis. For example, the work of ~\cite{13esteva2017dermatologist} trained a VGG model for skin cancer diagnosis using the photos captured with a daily camera. 

The deep learning method has been also utilized in \cite{14meng2017liver} for the same task studied in our paper. Its algorithm for liver fibrosis diagnosis is formulated as a classification task that takes only one ultrasound image (usually liver parenchyma echo) as input. However, this method does not resemble the practical clinician's process, which usually utilizes multiple ultrasound images for diagnosis. Due to insufficient input information, it tends to result in low prediction accuracy. 

The clinical workflow of liver fibrosis is consisting of two primary steps: Firstly, the doctors usually scan 10 ultrasound images for different body locations and then predict a set of indicators. Secondly, they conduct the diagnosis by observing both the indicators and the scanned images. Our approach is designed following this procedure from two aspects: 1) A multi-stream feature modeling module is developed to extract features from the 10 images in parallel, which are then concatenated for further liver fibrosis classification. This is implemented by a VGG model with weights sharing. 2) We innovatively involve the multi-indicator labels as extra supervisions and form an indicator-guided learning scheme. The indicators are connected with their corresponding feature streams based on the clinical guidelines. These strategies not only efficiently improve the prediction accuracy but also make the automatic diagnosis interpretable. To support the training, a dataset is carefully collected, which contains the samples of 229 patients. For each patient, 8 ultrasound videos and 10 ultrasound images are scanned, the results of 13 indicators and the final diagnosis are also collected. 

To this end, the contributions of this paper can be concluded as: 
\begin{itemize}
	\item An novel algorithm for automatic liver fibrosis diagnosis is proposed, which well-follows the clinical workflow and achieves the state-of-the-art performance.
	\item A dataset is carefully-prepared, including the ultrasound videos/images and the results of both 13 indicators and the final diagnosis.  
	\item A novel indicator-guided multi-stream deep neural network is designed which efficiently ease the training procedure and also makes the model interpretable.  
\end{itemize}



\section{Dataset Construction}
According to the radiologists' routine diagnostic procedure, by only exploiting one ultrasound image containing liver echo information provides insufficient information for the liver fibrosis diagnosis. Thus, we designed a protocol to collect ultrasound dataset with multiple ultrasound images and videos, aiming at promoting the development of automatic diagnosis of liver fibrosis.

\subsection{Clinical workflow}
The diagnosis process for radiologists is very instructive. Specifically, radiologists first saved 10 indicators images for each patient. Then 8 videos are also saved for supplemented temporal information. We follow the same process to collect our own dataset.

The concrete diagnosis process is described below. First, place the ultrasound probe on the left side of the xiphoid and  obtain a clear image of the left hepatic angle by vertical section scanning. Second, move to the place below the right costal margin, we will see right liver and hepatic vein. Third, move to the place below the left costal margin, and there are spleen. Then, we will place the probe on right intercostal space, and obtain the image of portal vein. And we can obtain the image of liver parenchyma  concatenated with the image of spleen. Based on domain knowledge, liver parenchyma echo should be distinguished with spleen parenchyma echo. Except the image of liver capsule is obtained through convex array probe, the others are obtained by linear array probe.

\subsection{Dataset details}

\paragraph{{\textbf{Patients}}}
Since our research focuses on liver disease caused by hepatitis B, patients collected in our dataset are primarily infected with hepatitis B. Besides, those who have received liver surgery, gallbladder surgery, or spleen surgery are further excluded as information of these organ will affect liver fibrosis diagnosis. In total, we collect the samples for 229 patients from a hospital. 

\begin{figure}[htbp] 
\centering 
\includegraphics[height=2.5cm, width=10cm]{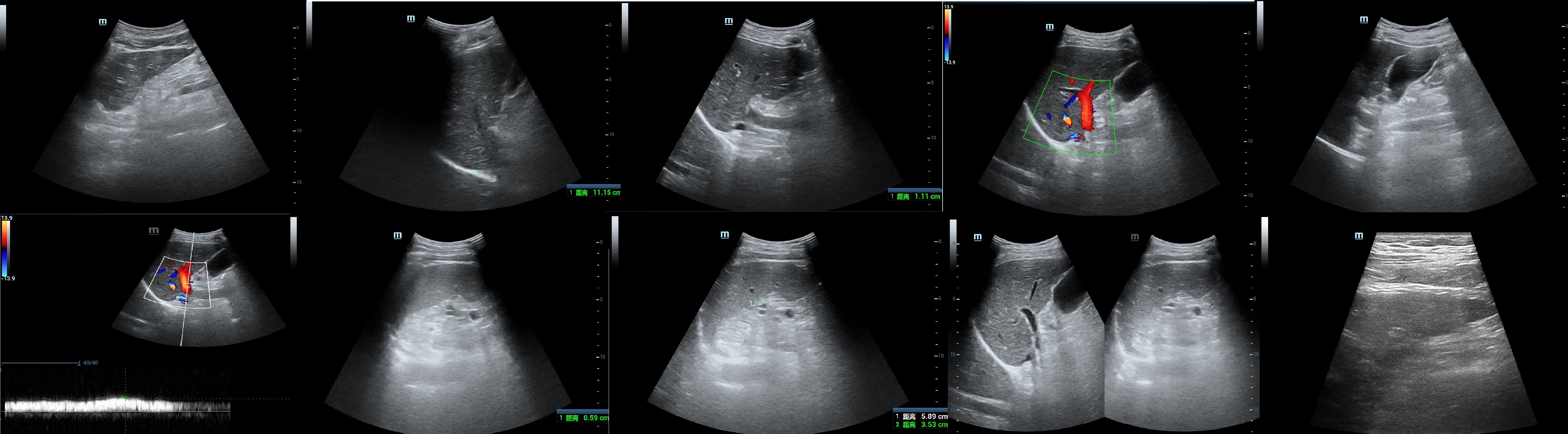} 
\caption{Indicator Images} \label{fig1}
\end{figure}

\paragraph{{\textbf{Indicators}}}
Before conducting final diagnosis, the clinical doctors usually predict a set of indicators by observing the statuses of 10 ultrasound images in different body locations. We call these images as indicator images (illustrated in Fig ~\ref{fig1}). 
Due to the lack of authoritative ultrasound diagnosis guideline for liver fibrosis, we investigate relevant medical literature~\cite{mediccrespo2012arfi} and consult with experienced radiologists, concluding 13 indicators, i.e., left hepatic angle, right liver slant diameter, gallbladder wall morphology, gallbladder wall thickness, spleen size, spleen thickness, spleen Length, liver parenchyma echo, liver capsule morphology, portal vein diameter, portal vein blood flow direction, hepatic vein diameter and hepatic vein morphology. Each indicator is of multiple statues, making the indicator prediction can be formulated as a classification task. For example, on the basis of indicator 1, left hepatic angle of a patient can have two categories, i.e., acute or blunt. The classification labels of those indicators are given by experienced radiologists.


\paragraph{\textbf{Indicator images}}
The ultrasound images containing the indicators information are called indicator images. In consistent with practical clinical procedure, we also collect 10 indicator images for each patient for 13 indicators judgement. Note that, the 10 indicator images and the 13 indicators have a many-to-many mapping. 
This is given by experience radiologists and is very helpful for our model design. For example, indicator image1 contains information about indicator 1, i.e., left hepatic angle.

\paragraph{\textbf{Ultrasound videos}}
In order to supplement the temporal coherence information of indicators that the radiologists may use during the diagnosis, each patient collected 8 ultrasound videos and each video lasted 5 seconds. For example, the video\_4 corresponding indicator\_6 which is the liver capsule morphology.

\paragraph{\textbf{Diagnoses}}
Diagnosis is given based on the information of all indicators. All patients were divided into 4 categories by experienced radiologists, which are normal, coarseness of liver parenchyma echo, liver ﬁbrosis and liver cirrhosis.

\paragraph{\textbf{Dataset summary}}
All the dataset summaries are listed in Table~\ref{tab:Data summary}, which consists of distribution of diagnoses and indicators for 229 patients. 
\begin{table}[]
\caption{Data summary}
\resizebox{\textwidth}{!}{
\begin{tabular}{|c|c|c|c|c|c|c|c|}

\hline
\rowcolor[HTML]{C0C0C0} 
Label                     & Diagnoses                                                                      & \begin{tabular}[c]{@{}c@{}}Indicator\_1 \\ (Left hepatic angle)\end{tabular}         & \begin{tabular}[c]{@{}c@{}}Indicator\_2\\  (Liver size)\end{tabular}              & \begin{tabular}[c]{@{}c@{}}Indicator\_3 \\ (Right liver slant)\end{tabular} & \begin{tabular}[c]{@{}c@{}}Indicator\_4\\  (Liver parenchyma echo)\end{tabular} & \begin{tabular}[c]{@{}c@{}}Indicator\_5 \\ (Spleen size)\end{tabular}                & \begin{tabular}[c]{@{}c@{}}Indicator\_6\\  (Liver capsule form)\end{tabular}                \\ \hline
\cellcolor[HTML]{FFFE65}0 & \begin{tabular}[c]{@{}c@{}}38\\  (Normal)\end{tabular}                         & \begin{tabular}[c]{@{}c@{}}149 \\ (Acute )\end{tabular}                         & \begin{tabular}[c]{@{}c@{}}180 \\ (Normal)\end{tabular}                           & \begin{tabular}[c]{@{}c@{}}124 \\ (Less than 130mm)\end{tabular}            & \begin{tabular}[c]{@{}c@{}}138\\ (coarseness of liver parenchyma echo)\end{tabular}                 & \begin{tabular}[c]{@{}c@{}}174 \\ (Mild swelling)\end{tabular}                        & \begin{tabular}[c]{@{}c@{}}122\\  (Smooth)\end{tabular}                                \\ \hline
\cellcolor[HTML]{FFFE65}1 & \begin{tabular}[c]{@{}c@{}}73\\ (Coarseness of liver parenchyma echo)\end{tabular}                      & \begin{tabular}[c]{@{}c@{}}80 \\ (Blunt)\end{tabular}                          & \begin{tabular}[c]{@{}c@{}}49\\  (Zoom out)\end{tabular}                          & \begin{tabular}[c]{@{}c@{}}105 \\ (Larger than 130mm)\end{tabular}          & \begin{tabular}[c]{@{}c@{}}12\\ (Asymmetry)\end{tabular}                   & \begin{tabular}[c]{@{}c@{}}22 \\ (Moderate swelling)\end{tabular}                     & \begin{tabular}[c]{@{}c@{}}63 \\ (Wavy)\end{tabular}                                   \\ \hline
\cellcolor[HTML]{FFFE65}2 & \begin{tabular}[c]{@{}c@{}}58\\ (Liver Fibrosis)\end{tabular}                  & -                                                                                    & -                                                                                 & -                                                                           & \begin{tabular}[c]{@{}c@{}}58\\ (Patch)\end{tabular}                       & 19 (Severe swelling)                                                                  & \begin{tabular}[c]{@{}c@{}}44 \\ (Jagged)\end{tabular}                                 \\ \hline
\cellcolor[HTML]{FFFE65}3 & \begin{tabular}[c]{@{}c@{}}60\\ (Liver Cirrhosis)\end{tabular}                 & -                                                                                    & -                                                                                 & -                                                                           & -                                                                          & -                                                                                     & -                                                                                      \\ \hline
\rowcolor[HTML]{C0C0C0} 
Label                     & \begin{tabular}[c]{@{}c@{}}Indicator\_7 \\ (Portal vein diameter)\end{tabular} & \begin{tabular}[c]{@{}c@{}}Indicator\_8\\  (Portal vein flow direction)\end{tabular} & \begin{tabular}[c]{@{}c@{}}Indicator\_9\\  (Hepatic vein morphology)\end{tabular} & \begin{tabular}[c]{@{}c@{}}Indicator\_10\\  (Spleen thickness)\end{tabular}     & \begin{tabular}[c]{@{}c@{}}Indicator\_11\\  (Spleen length)\end{tabular}   & \begin{tabular}[c]{@{}c@{}}Indicator\_12 \\ (Gallbladder wall thickness)\end{tabular} & \begin{tabular}[c]{@{}c@{}}Indicator\_13 \\ (Gallbladder wall morphology)\end{tabular} \\ \hline
\cellcolor[HTML]{FFFE65}0 & \begin{tabular}[c]{@{}c@{}}146 \\ (Less than 12mm)\end{tabular}                & \begin{tabular}[c]{@{}c@{}}224  \\ (Into the liver)\end{tabular}                     & \begin{tabular}[c]{@{}c@{}}185 \\ (Stiffness)\end{tabular}                        & \begin{tabular}[c]{@{}c@{}}160 \\ (Less than 40mm)\end{tabular}             & \begin{tabular}[c]{@{}c@{}}167 \\ (Less than 120mm)\end{tabular}           & \begin{tabular}[c]{@{}c@{}}214 \\ (Less than 3mm)\end{tabular}                        & \begin{tabular}[c]{@{}c@{}}49 \\ (Smooth)\end{tabular}                                 \\ \hline
\cellcolor[HTML]{FFFE65}1 & \begin{tabular}[c]{@{}c@{}}83 \\ (Larger than 12mm)\end{tabular}               & \begin{tabular}[c]{@{}c@{}}5 \\ (Out the liver)\end{tabular}                         & \begin{tabular}[c]{@{}c@{}}44 \\ (Slim)\end{tabular}                              & \begin{tabular}[c]{@{}c@{}}55\\  (Larger than 40mm)\end{tabular}            & \begin{tabular}[c]{@{}c@{}}48 \\ (Less than 120mm)\end{tabular}            & \begin{tabular}[c]{@{}c@{}}8 \\ (Larger than 3mm)\end{tabular}                        & \begin{tabular}[c]{@{}c@{}}173 \\ (Rough)\end{tabular}                                 \\ \hline
\end{tabular}}
\label{tab:Data summary}
\end{table}

\section{Methodology}

\subsection{Multi-stream Feature Extraction}
As illustrated in Fig.~\ref{fig:F1}, 10 indicator images are fed into 10 VGG models (a vanilla VGG-16 without last two fully connected layers) in parallel, which achieves a 4096-D feature vector for each indicator image. To reduce the model complexity and enhance feature learning between different indicator images, we exploit shared weights for 10 parallel VGG models. Such weight-sharing strategy means the same weights are leveraged for the forward and the total losses are accumulated from 10 parallel paths for the backward, which can avoid over-fitting problem effectively.


\subsection{Indicator-guided Learning}
\begin{figure}[htbp] 
\centering 
\includegraphics[height=6cm, width=12cm]{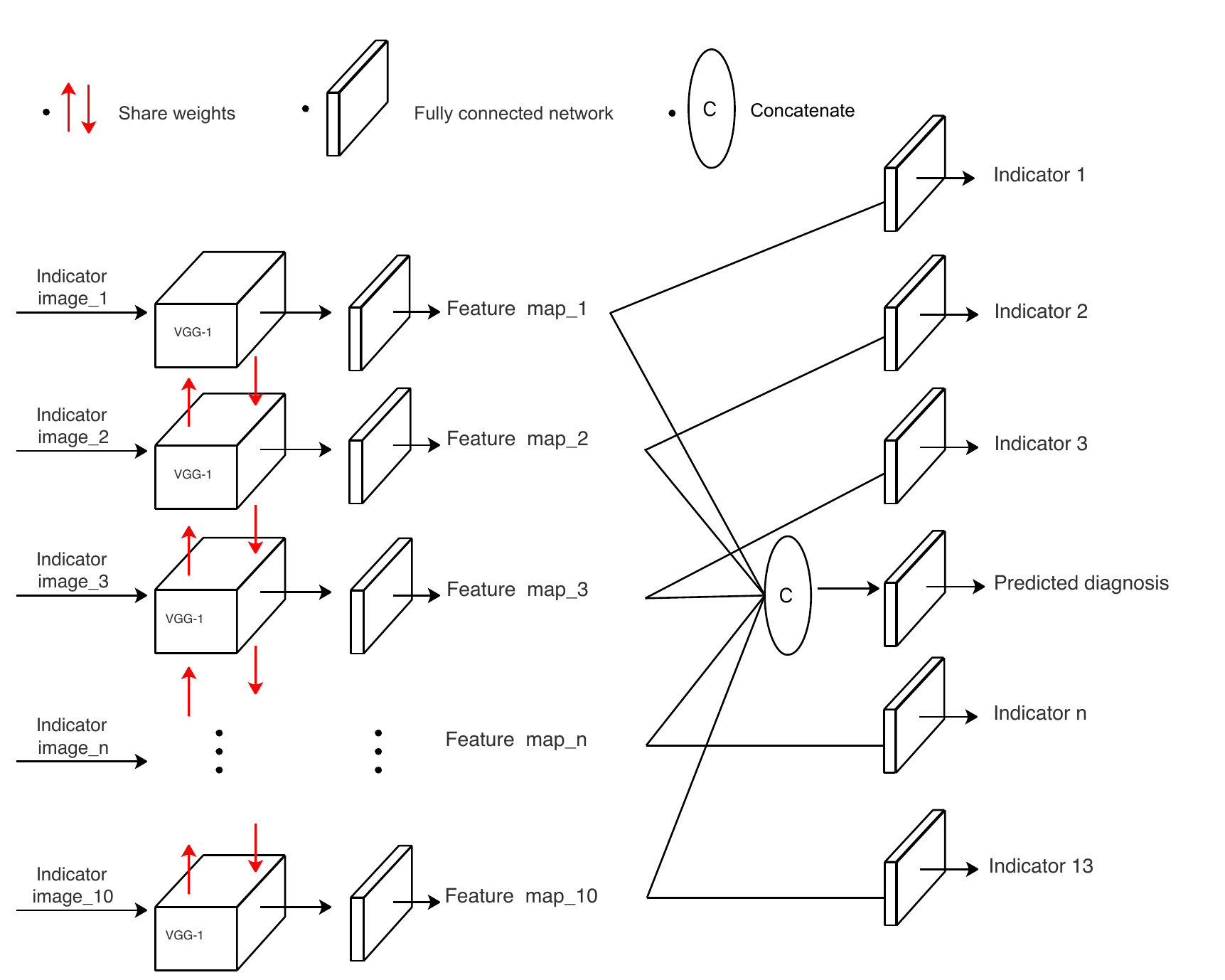} 
\caption{Multi-indicator guided deep neural networks for ultrasound liver fibrosis diagnosis with weight-sharing.} \label{fig:F1}
\end{figure}
As the whole pipeline shown in Fig.~\ref{fig:F1}, then the feature map of one indicator image extracted in multi-stream feature extraction module is divided to two streams. One is fed to a fully connected layer to predict corresponding indicator label. The other one is concatenated with other 9 indicator image's feature maps as the patient level feature to predict diagnosis label through another fully connected layer. 

\paragraph{\textbf{Loss functions.}} We design two losses to train our model, i.e., Indicator Loss and Diagnosis Loss. Indicator Loss and Diagnosis Loss are the cross-entropy losses between predicted labels and their ground truth labels. We use the following weighted sum of $n$ Indicator Loss and Diagnosis Loss as our Total Loss for joint learning.


$$Total\_Loss = Diagnosis\_Loss + \sum_{i=0}^n \lambda_i * Indicator\_Loss_i$$

\subsection{Training Strategy}
In order to get good performance, we utilize several training strategies as follows.

\paragraph{\textbf{Pre-training using ultrasound videos}}  Vanilla VGG is a pre-trained model on ImageNet dataset consisting of nature images, so it cannot be used as feature extractor for ultrasound images directly. Thus, we first fine-tune the VGG model based on the large amount of frames extracted from ultrasound videos, detailed setting is presented in Section~\ref{ID}.

\paragraph{\textbf{3-Stage Optimization}}

In training stage, our model aims to predict indicator label and diagnosis label, which is known as a multi-task learning task. To accelerate model training, we utilize a 3-stage-training strategy. In the first stage, we focus on learning indicators, i.e., we fix the parameters of diagnosis prediction layers and only use Indicator Loss to update our model. Corresponding to the first stage, we fix the parameters of indicator prediction layers and use Diagnosis Loss to update our model. In final stage, we use the joint Total Loss as introduced above to optimize the whole model.

\section{Experiments}

\subsection{Implementation Details}
\label{ID}
Our model is implemented in Pytorch\footnote{\url{https://pytorch.org/}}. We use video frames to pre-train our model. Each video is sampled one frame per second. And the label of each frame is assigned the same as the corresponding video. Based on the pre-trained model, we further fine-tune our model using indicator images. In each batch, $10$ indicator images from one patient as inputs are fed into the model, and the patient's diagnosis label and indicator labels work as supervision. We use Momentum SGD with weight decay as optimization method. The initial learning rate is $2 * 10^{-3}$ and momentum is set as $0.9$.We trained our model by 50 epochs with non-decreasing model policy. For data augmentation, we use random resize crop and random flip. And we set $\lambda = 0.1$ experimentally.

\begin{table}[h]
  \caption{Experimental Results}
  \centering
  \begin{tabular}{cc}
    \toprule
    \cmidrule{1-2}
    Methods   &   Accuracy     \\
    \midrule
    \cite{14meng2017liver}   & 45.6\% \\
    \textbf{Ours} & \textbf{65.6}\%  \\
    Without 3-stage  & 59.5\%   \\
    Without Indicator  & 56.2\%   \\
    Without Weight Sharing   & 55.0\%  \\
    Without Pre-train  & 40.6\%   \\
    \bottomrule
  \end{tabular}  \label{tab:results}
\end{table}

\subsection{Comparison against State-of-the-Arts}
To the best of our knowledge, only~\cite{14meng2017liver} designs a deep learning framework for the liver fibrosis diagnosis using ultrasound images. Thus, we reproduce their method and compare with our methods on our large dataset. As the results presented in Table \ref{tab:results}, Our method outperforms ~\cite{14meng2017liver} by a large margin.

\subsection{Ablation Studies}

To discover the vital elements or components of our model, we conduct an ablation study by removing or reap lacing some components. The detailed experimental results are also shown in Table \ref{tab:results}.

{\textbf{w/o indicator guidance}} In order to evaluate the effect of indicator guidance, we simplify our model by removing the indicator supervision and just use diagnosis labels for the training. The results show that our model can get better performance with an extra supervision of indicator labels.

{\textbf{w/o video pre-training}} Without using ultrasound data to pre-train VGG, the diagnosis accuracy decreases dramatically, which shows the effectiveness of pre-training.

{\textbf{w/o weights sharing}} Rather than using a shared weight VGG to extract feature from 10 kinds of indicator images, we use 10 different VGG models. The results show that weight sharing is a good way to reduce over-fitting, which is essential for small dataset.

{\textbf{w/o 3-Stage Optimization}} Rather than using three stages training strategy, we compare results training with only one stage, which exploits the combined Diagnosis Loss and Indicator Loss to train the whole model jointly. The results show that one stage learning is not stable to get the optimal solution.

\section{Conclusion}
In this paper, a deep learning framework is presented for automatically liver fibrosis prediction. Our approach can take using the information provided by multiple ultrasound images. An indicator-guided learning mechanism is further proposed to ease the training of the proposed network through weight-sharing. This follows the workflow of clinical diagnosis and make the prediction procedure interpretable. Besides, a dataset is well-collected which contains the ultrasound videos/images, indicators and labels of 229 patients. Our proposed model shows its effectiveness by achieving the state-of-the-art performance. In the future, how to fully encode the information from ultrasound video instead of selected frames and how to fuse video and indicator images information better may be two possible research directions. 

%
%
%
\bibliography{main.bib}
\bibliographystyle{unsrt}

\end{document}